\documentstyle[12pt]{article}
\begin{document}

\begin{titlepage}

\begin{center}{\LARGE
Nonlocality, Self-Adjointness and $\Theta$-Vacuum in Quantum Field Theory
in Spaces with Nontrivial Topology}
\end{center}
\vskip 3 cm
\begin{center}
{\large Yu.A.~Sitenko}\\
\mbox{} \\
{\it Bogolyubov Institute for Theoretical Physics, 
National Academy of Sciences of Ukraine,} \\
{\it  252143 Kiev, Ukraine}
\end{center}

\vskip 1 cm
\begin{abstract}
We consider an analogue of the Aharonov-Bohm effect in quantum field theory:
the fermionic vacuum attains nontrivial quantum numbers in the background of
a magnetic vortex even in the case when the spatial region of nonvanishing
external field strength is excluded.  The dependence of the vacuum quantum
numbers on the value of the vortex flux and the choice of the condition on 
the boundary of the excluded region is determined.
\end{abstract}

\medskip

\end{titlepage}

\setcounter{page}{1}

As is known, the singular static magnetic monopole background induces
fermion number in the vacuum [1-3]
\begin{equation}
\langle N\rangle=-{1\over\pi}\arctan\left(\tan\frac{\Theta}{2}\right),
\end{equation}
where $\Theta$ is the parameter of a self-adjoint extension, which defines 
the boundary condition at a puncture corresponding to the location of the 
monopole; this results in the monopole becoming actually the dyon violating 
the Dirac quantization condition and CP symmetry.

In the present talk I shall be considering quantum numbers which are 
induced in the fermionic vacuum by the singular static magnetic string 
background.
Since
the deletion of a line , as compared to the deletion of a point, changes the
topology of space in a much more essential way (fundamental group becomes
nontrivial), the properties of the $\Theta$-vacuum will appear to be much 
more diverse, as compared to Eq.(1).  
Restricting ourselves to a surface which is
orthogonal to the string axis, let us consider 2+1-dimentional spinor
electrodynamics on a plane with a puncture corresponding to the location
of the string.  I shall show that in this case the induced vacuum fermion
number  and magnetic flux depend on the self-adjoint extension parameter
and the magnetic flux of the string as well.

The pertinent Dirac Hamiltonian has the form
\begin{equation}
H=-i\vec{\alpha}[\vec{\partial}-i\vec{V}(\vec{x})]+\beta m;
\end{equation}
where $\vec{V}(\vec{x})$ is an external static vector potential.  In a flat
two-dimensional space ($\vec{x}=(x^1,x^2)$) the vacuum fermion number 
induced
by such a background was calculated first in Ref.[4]
\begin{equation}
\langle N\rangle=-{1\over2}{\rm sgn}(m)\Phi,
\end{equation}
where ${\rm sgn}(u)=\left\{\begin{array}{rl}1,&u>0\\
-1,&u<0\end{array}\right.$
is the
sign function and $\Phi={1\over2\pi}\int d^2x B(\vec{x})$ is
the total flux (in
the units of $2\pi$) of the external magnetic field strength
$B(\vec{x})=\vec{\partial}\times \vec{V}(\vec{x})$ piercing the
two-dimensional space (plane); note that the mass parameter $m$ in
Eq.(2) can take both positive and negative values in two and any even
number of spatial dimensions.

It should be emphasized, however, that Eq.(3) is valid for  regular
external field configurations only, i.e. $B(\vec{x})= B_{\rm
reg}(\vec{x})$, where $B_{\rm reg}(\vec{x})$  is a
continuous in the whole function that can grow at most as
$O\bigl(|\vec{x}-\vec{x}_s|^{-2+\varepsilon}\bigr)$ $(\varepsilon>0)$
at separate points; as to a vector potential
$\vec{V}(\vec{x})=\bigl(V_1(\vec{x}),V_2(\vec{x})\bigr)$, it is
unambiguously defined everywhere on the plane. The regular
configuration of an external field polarizes the vacuum locally, and
Eq.(3) is just the integrated version of the linear relation between
the vacuum fermion number density and the magnetic field strength.

One can ask the following question: whether the nonlocal effects of
the external field background are possible, i.e., if the spatial
region of nonvanishing field strength is excluded, whether there will
be vacuum polarization in the remaining part of space? For the
positive answer it is necessary, although not sufficient, that the
latter spatial region be of nontrivial topology [5] (see
also Ref.[6]). However, the condition on the boundary of
the excluded region has not been completely specified. In the present
talk this point will be clarified by considering the whole set of
boundary conditions which are compatible with the self-adjointness
of the Dirac Hamiltonian in the remaining region.

We shall be interested in the situation when the volume of the
excluded region is shrinked to zero, while the global characteristics
of the external field in the excluded region is retained nonvanishing.
This implies that singular, as well as regular, configurations of
external fields have to be considered. In particular, in two spatial
dimensions the magnetic field strength is taken to be a distribution
(generalized function)
\begin{equation}
B(\vec{x})=B_{\rm reg}(\vec{x})+2\pi\Phi^{(0)}\delta(\vec{x}),
\end{equation}
where $\Phi^{(0)}$ is the total magnetic flux (in the units of $2\pi$)
in the excluded region which is placed at the origin $\vec{x}=0$. As
to the vector potential, it is unambiguously defined everywhere with
the exception of the origin, i.e. the limiting value
$\lim_{|\vec{x}|\to0}\vec{V}(\vec{x})$ does not exist, or, to be more
precise, a singular magnetic vortex is located at the origin
\begin{equation}
\lim_{|\vec{x}|\to0}\vec{x}\times\vec{V}(\vec{x})=\Phi^{(0)}.
\end{equation}
Certainly, a plane has trivial topology, $\pi_1=0$, while a plane with
a puncture where the vortex is located has nontrivial topology,
$\pi_1={{\rm Z}}$; here ${{\rm Z}}$ is the set of integer numbers and
$\pi_1$ is the first homotopy group of the surface.

The total magnetic flux through the punctured plane is obviously
defined as
\begin{equation}
\Phi={1\over2\pi}\int d^2x B_{\rm
reg}(\vec{x})={1\over2\pi}\int\limits_0^{2\pi}d\varphi
\bigl[\vec{x}\times\vec{V}(\vec{x})\bigr]\biggm|^{r=\infty}_{r=0},
\end{equation}
where the polar coordinates $r=|\vec{x}|$ and
$\varphi=\arctan(x^2/x^1)$ are introduced.

The Dirac equation with the Hamiltonian (2) on a punctured plane is
invariant with respect to the gauge transformations
\begin{equation}
G: \vec{V}(\vec{x})\to\vec{V}(\vec{x})+\vec{\partial}\Lambda(\vec{x}),
\quad\psi(\vec{x})\to{\rm e}^{i\Lambda(\vec{x})}\psi(\vec{x}).
\end{equation}
Although the vector potential in any gauge is single-valued on a
punctured plane, this is not true for the gauge function
$\Lambda(\vec{x})$. Since the magnetic flux $\Phi$(6) (and the field
strength $B_{\rm reg}(\vec{x})$) remains invariant under gauge
transformations, the most general condition on  $\Lambda(\vec{x})$
takes the form
\begin{equation}
\Lambda(r,\varphi+2\pi)=\Lambda(r,\varphi)+2\pi\Upsilon_\Lambda,
\end{equation}
where $\Upsilon_\Lambda$ is the independent of $r$ and $\varphi$
parameter of the gauge transformation; incidentally the magnetic flux
of the vortex $\Phi^{(0)}$(5) is changed: $\Phi^{(0)}\to\Phi^{(0)}+
\Upsilon_\Lambda$. If one takes a single-valued wave function,
$\psi(r,\varphi+2\pi)=\psi(r,\varphi)$, then, after applying a gauge
transformation to it, one gets a wave function satisfying the condition
$({\rm e}^{i\Lambda}\psi)(r,\varphi+2\pi)={\rm
e}^{i2\pi\Upsilon_\Lambda}({\rm e}^{i\Lambda}\psi)(r,\varphi)$. Thus
the set of wave functions on a punctured plane is much richer than
that of wave functions on a plane without a puncture (in the latter
case only the gauge transformations with $\Upsilon_\Lambda=0$ are
admissible). Certainly, there are no reasons to impose the condition
of single-valuedness on the initial function, and in the most general
case one takes
\begin{equation}
\psi(r,\varphi+2\pi)={\rm
e}^{i2\pi\Upsilon}\psi(r,\varphi), \end{equation}
and after applying a gauge transformation one gets \begin{equation}
({\rm e}^{i\Lambda}\psi)(r,\varphi+2\pi)={\rm
e}^{i2\pi(\Upsilon+\Upsilon_\Lambda)}({\rm
e}^{i\Lambda}\psi)(r,\varphi).
\end{equation}
Therefore, if one admits singular gauge transformations
$(\Upsilon_\Lambda\neq0)$, as well as regular ones
$(\Upsilon_\Lambda=0)$, then one has to consider wave functions
defined on a plane with a cut which starts from the puncture and goes
to infinity in the radial direction at, say, the angle
$\varphi=\varphi_c$. The boundary conditions on the sides of the cut
are globally parametrized by the values of $\Upsilon$.

All this can be presented in a more refined way, using the notion of a
self-adjoint extension of a Hermitian (symmetric) operator. The
orbital angular momentum operator, $-i\partial_\varphi$, entering the
Dirac Hamiltonian(2) is Hermitian, but not self-adjoint, when defined
on the domain of functions satisfying, say,
$\psi(r,\varphi_c+2\pi)=\psi(r,\varphi_c)=0$; this operator has the
deficiency index equal to (1,1). The use of the Weyl--von Neumann
theory of self-adjoint extension [7] yields that
$-i\partial_\varphi$ becomes self-adjoint, when defined on the domain
of functions satisfying Eq.(9) with $\varphi=\varphi_c$, where the
values of $\Upsilon$ parametrize the family of extensions. It should
be stressed that    $\Upsilon$, as well as $\Phi^{(0)}$, is changed
under the singular gauge transformations (compare Eqs. (9) and (10)),
while the difference  $\Phi^{(0)}-\Upsilon$ remains invariant.

Let us turn now to the boundary condition at the puncture $\vec{x}=0$.
In the following our concern will be in the case in which the regular part
of the magnetic field is absent, $B_{\rm reg}(\vec{x})=0$. Then, in
the representation with $\alpha_1=\sigma_1$,  $\alpha_2=\sigma_2$ and
$\beta=\sigma_3$ ($\sigma_j$ are the Pauli matrices) the spinor wave
function satisfying the Dirac equation and the condition (9) has the
form
\begin{equation}
\psi(\vec{x})=\sum_{n\in{\rm Z}}\left(\begin{array}{l}f_n(r)
\exp[i(n+\Upsilon)
\varphi]\\g_n(r)\exp[i(n+1+
\Upsilon)\varphi]\end{array}\right),
\end{equation}
where the radial functions, in general, are
\begin{equation}
\left(\begin{array}{l}f_n(r)\\g_n(r)\end{array}\right)=
\left(\begin{array}{c}C^{(1)}_n(E)J_{n-\Phi^{(0)}+
\Upsilon}(kr)+C^{(2)}_n(E)Y_{n-\Phi^{(0)}+\Upsilon}(kr)\\
{ik\over E+m}\bigl[C^{(1)}_n(E)J_{n+1-\Phi^{(0)}+
\Upsilon}(kr)+C^{(2)}_n(E)Y_{n+1-\Phi^{(0)}+\Upsilon}(kr)\bigr]
\end{array}\right),
\end{equation}
$k=\sqrt{E^2-m^2}$, $J_\mu(z)$ and $Y_\mu(z)$ are the Bessel and the
Neumann functions of the order $\mu$. It is clear that the condition
of regularity at $r=0$ can be imposed on both $f_n$ and $g_n$ for all
$n$ in the case of integer values of the quantity
$\Phi^{(0)}-\Upsilon$ only. Otherwise, the condition of regularity at
$r=0$ can be imposed on both $f_n$ and $g_n$ for all but $n=n_0$,
where
\begin{equation}
n_0={[\![}\Phi^{(0)}-\Upsilon{]\!]}\;,
\end{equation}
${[\![} u{]\!]}$ is the integer part of the quantity $u$ (i.e. the
integer which is less than or equal to $u$); in this case at least one
of the functions, $f_{n_0}$ or $g_{n_0}$, remains irregular, although
square integrable, with the asymptotics $r^{-p}$ $(p<1)$ at $r\to0$
[8].  The question arises then, what boundary condition, instead of
regularity, is to be imposed on $f_{n_0}$ and $g_{n_0}$ at $r=0$ in
the latter case?

To answer this question, one has to find the self-adjoint extension
for the partial Hamiltonian  corresponding to the mode with $n=n_0$.
If this Hamiltonian is defined on the domain of regular at $r=0$
functions, then it is Hermitian, but not self-adjoint, having the
deficiency index equal to (1,1). Hence the family of self-adjoint
extensions is labeled by one real continuous parameter denoted in the
following by $\Theta$. It can be shown (see Ref.[9]) that,
for the partial Hamiltonian to be self-adjoint, it has to be defined
on the domain of functions satisfying the boundary condition 
\begin{equation}
\lim_{r\to0}\cos\biggl({\Theta\over2}+{\pi\over4}\biggr)\biggl
(|m|r\biggr)^Ff_{n_0}(r)=i\lim_{r\to0}\sin\biggl({\Theta\over2}+{\pi\over4}
\biggr)\biggl(|m|r\biggr)^{1-F}g_{n_0}(r),
\end{equation}
where
\begin{equation}
F={\{\hspace{-3.3pt}|}\Phi^{(0)}-\Upsilon{\}\!\!\!|}\;,
\end{equation}
${\{\hspace{-3.3pt}|} u{\}\!\!\!|}$ is the fractional part of the quantity 
$u$,
${\{\hspace{-3.3pt}|} u{\}\!\!\!|}=u-{[\![} u{]\!]}$, 
$0\leq{\{\hspace{-3.3pt}|} u{\}\!\!\!|}<1$; note here that
Eq.(14) implies that $0<F<1$, since in the case of $F=0$ both
$f_{n_0}$ and $g_{n_0}$ satisfy the condition of regularity at $r=0$.

Using the explicit form of the solution to the Dirac equation in the
background of a singular magnetic vortex, it is straightforward to
calculate the vacuum fermion number induced on a punctured plane.
As follows already from the preceding discussion, the vacuum fermion
number vanishes in the case of integer values of $\Phi^{(0)}-\Upsilon$
$(F=0)$, since this case is indistinguishable from the case of the trivial
background, $\Phi^{(0)}=\Upsilon=0$. In the case of noninteger values
of $\Phi^{(0)}-\Upsilon$ ($0<F<1$)
we get (details will be published elsewhere)
\begin{eqnarray}
\langle N\rangle=\left\{
\begin{array}{l}
\displaystyle{-\frac{1}{2}{\rm sgn}(m)F\;, \qquad
\Theta=\frac{\pi}{2} ({\rm mod} 2\pi)} \\[5mm]
\displaystyle{\frac{1}{2}{\rm sgn}(m)\left(1-F\right)\;, \qquad
\Theta=-\frac{\pi}{2} ({\rm mod} 2\pi)} \\
\end{array}
\right.
\end{eqnarray}
and
\begin{eqnarray}
& &\langle N\rangle=-{1\over2}
{\rm sgn}(m)\biggl(F-{1\over2}\biggr)-
{1\over4\pi}\int\limits_1^\infty
{dv\over v\sqrt{v-1}}\times \nonumber \\
& & \times{{\rm
sgn}(m)(Av^F-A^{-1}v^{1-F})+4\bigl(F-{1\over2}\bigr)\bigl(v-1\bigr)\over
Av^F+2{\rm sgn}(m)+A^{-1}v^{1-F}},\quad
\Theta \neq \frac{\pi}{2}({\rm mod} \pi),
%\nonumber \\
%& &\hspace*{1cm} \Theta \neq \frac{\pi}{2}({\rm mod} \pi),
\end{eqnarray}
where
\begin{equation}
A=2^{1-2F}{\Gamma(1-F)\over\Gamma(F)}\tan
\biggl({\Theta\over2}+{\pi\over4}\biggr),
\end{equation}
$\Gamma(u)$ is the Euler gamma-function.
Eqs.(16) and (17) can be presented in another form
\begin{eqnarray}
& &\langle N\rangle=\left\{
\begin{array}{ll}
-\frac{1}{2}{\rm sgn}(m)F +& \\[5mm]
+\frac{1}{2}
{\rm sgn}\left[A+{\rm sgn}(m)\right],
&\Theta \neq \frac{\pi}{2}({\rm mod} 2\pi), \\[5mm]
-\frac{1}{2}{\rm sgn}(m)F,
&\Theta = \frac{\pi}{2}({\rm mod} 2\pi),
\end{array}
\right\},\; \hspace*{5mm}
0<F<\frac{1}{2}\;, \\[5mm]
%\end{eqnarray}
%\begin{equation}
& &\langle N\rangle=-{1\over\pi}\arctan\biggl\{\tan\biggl[{\Theta\over2}+
{\pi\over4}\bigl(1-{\rm sgn}(m)\bigr)\biggr]\biggr\}\quad,
\hspace*{15mm} F={1\over2}\;, \qquad \\[5mm]
%\end{equation}
%\begin{eqnarray}
& &\langle N\rangle=\left\{
\begin{array}{ll}
\frac{1}{2}{\rm sgn}(m)\left(1-F\right)- &\\[5mm]
-\frac{1}{2}
{\rm sgn}\left[A^{-1}+{\rm sgn}(m)\right],
&\Theta \neq -\frac{\pi}{2}({\rm mod} 2\pi), \\[5mm]
\frac{1}{2}{\rm sgn}(m)\left(1-F\right),
&\Theta = -\frac{\pi}{2}({\rm mod} 2\pi),
\end{array}
\right\},\;
\frac{1}{2}<F<1\;;
\end{eqnarray}
note that Eq.(20) in the case of $m>0$, when it coincides with Eq.(1),
has been obtained earlier in Ref.[10]. We get also the relations
\begin{equation}
\lim_{F\to0}\langle N\rangle={1\over2}{\rm sgn}(m), \quad
\Theta \neq \frac{\pi}{2} ({\rm mod} 2\pi)
\end{equation}
and
\begin{equation}
\lim_{F\to1}\langle N\rangle=-{1\over2}{\rm sgn}(m), \quad
\Theta \neq -\frac{\pi}{2} ({\rm mod} 2\pi),
\end{equation}
indicating that the vacuum fermion number is not, in general,
continuous at integer values of $\Phi^{(0)}-\Upsilon$; the limiting
values (22) and (23) differ from the value at $F=0$ exactly, the latter
being equal, as noted before, to zero.

It is obvious that the vacuum fermion number at fixed values of
$\Upsilon$   and $\Theta$ is periodic in the value of
$\Phi^{(0)}$. This feature (periodicity in $\Phi^{(0)}$) is also
shared by the quantum-mechanical scattering of a nonrelativistic
particle in the background of a singular magnetic vortex, known as the
Aharonov-Bohm effect [11]. Since there appear assertions in
the literature which deny the periodicity of the vacuum fermion number
in $\Phi^{(0)}$ [12,13], the following comments on the
result (19)--(21) will be clarifying.

Under the charge conjugation,
\begin{equation}
C:\quad \vec{V}\to-\vec{V},\quad \psi\to\sigma_1\psi^*,\quad
\Upsilon\to-\Upsilon,
\end{equation}
the fermion number operator and its vacuum value are to be odd, $N\to
-N$ and $\langle N\rangle\to -\langle N\rangle$. Evidently, the result
(19)--(21) is not, since the boundary condition (14) breaks,
in general, the
charge conjugation symmetry. However, for certain choices of the
parameter $\Theta$ this symmetry can be retained [14].

In particular, choosing
\begin{equation}
\left.\begin{array}{ll}
\Theta={\pi\over2}({\rm mod}2\pi),&\Phi^{(0)}-\Upsilon>0\\
\Theta=-{\pi\over2}({\rm mod}2\pi),&\Phi^{(0)}-\Upsilon<0\end{array}
\right\}
\quad (\Phi^{(0)}-\Upsilon\neq n,\ \ n\in{{\rm Z}}),
\end{equation}
which corresponds to the boundary condition of
Refs.[15,16], one obtains [12, 13, 17]
\begin{equation}
\langle N\rangle= \left\{\begin{array}{ll} -{1\over2}{\rm
sgn}(m)F,&\Phi^{(0)}-\Upsilon>0\\ {1\over2}{\rm
sgn}(m)(1-F),&\Phi^{(0)}-\Upsilon<0\end{array}\right\},\quad 0<F<1,
\end{equation} which is odd under the charge conjugation but is not 
periodic in
$\Phi^{(0)}$.

No wonder that there exists a choice of $\Theta$ respecting both the
periodicity in $\Phi^{(0)}$ and the charge conjugation symmetry,
namely,
\begin{equation}\begin{array}{ll}
\Theta={\pi\over2}({\rm mod}2\pi),&0<F<{1\over2}\\
\Theta=-{\pi\over2}[1-{\rm sgn}(m)]({\rm mod}2\pi),&F={1\over2}\\
\Theta=-{\pi\over2}({\rm mod}2\pi),&{1\over2}<F<1\end{array},
\end{equation}
which corresponds to the condition of minimal irregularity, i.e. to
the radial functions being divergent at $r\to0$ at most as $r^{-p}$
with $p\leq{1\over2}$. This is the boundary condition, with the use of
which the result of Ref.[5] is obtained:
\begin{equation}
\langle N\rangle={1\over2}{\rm sgn}(m)\biggl[{1\over2}{\rm
sgn}_0\biggl(F-{1\over2}\biggr)-F+{1\over2}\biggr],
\end{equation}
where
$
{\rm sgn}_0(u)=\left\{
\begin{array}{ll}
{\rm sgn}(u),&u\neq0\\
0,&u=0\end{array}\right. .
$
Note that Eq.(28) is continuous at integer values of
$\Phi^{(0)}-\Upsilon$ and discontinuous at half-integer ones.

Another choice compatible with the periodicity in
$\Phi^{(0)}$ and the symmetry (24) is \linebreak $\Theta=-{\pi\over2}
[1-{\rm sgn}(m)]({\rm mod}2\pi)$ for $0<F<1$; then the vacuum fermion
number is discontinuous both at integer and half-integer values of
$\Phi^{(0)}-\Upsilon$.

We have calculated also the total magnetic flux induced in the
fermionic vacuum on a punctured plane
\begin{equation}
\Phi^{(I)}=-{e^2F(1-F)\over2\pi|m|}\left[{1\over6}\biggl(F-
{1\over2}\biggr)+
{1\over4\pi}\int\limits_1^\infty{dv\over
v\sqrt{v-1}}{Av^F-A^{-1}v^{1-F}\over Av^F+2{\rm
sgn}(m)+A^{-1}v^{1-F}}\right];
\end{equation}
note that the coupling constant $e$ relating the vacuum current to the
vacuum magnetic field strength (via the Maxwell equation) has the
dimension $\sqrt{|m|}$ in $2+1$-dimensional space-time. At
half-integer values of $\Phi^{(0)}-\Upsilon$ we get
\begin{equation}
\Phi^{(I)}=-{e^2\over8\pi^2m}\arctan\biggl\{\tan\biggl[{\Theta\over2}+
{\pi\over4}\bigl(1-{\rm sgn}(m)\bigr)\biggr]\biggr\}
\qquad\left(F=\frac{1}{2}\right).
\end{equation}
The vacuum magnetic flux under the boundary condition (25) is given in
Ref.[12]. Under the boundary condition (27) we get
\begin{equation}
\Phi^{(I)}={e^2F(1-F)\over12\pi|m|}\biggl[{3\over2}{\rm
sgn}_0\bigl(F-{1\over2}\bigr)-F+{1\over2}\biggr],
\end{equation}
which is both periodic in $\Phi^{(0)}$ and $C$-odd. As it follows from
Eq.(31), the vacuum under the boundary condition (27) is in a certain
sense of a diamagnetic type at $0 < F < \frac{1}{2}$ and of a
paramagnetic type at $\frac{1}{2} < F < 1$.

In conclusion, we present the evident relations
\begin{eqnarray}
& &\sum\limits_{{\rm sgn}(m)}\langle N\rangle=\left\{
\begin{array}{ll}
\frac{1}{2}
{\rm sgn}\left(A+1\right)+ & \\[2mm]
+\frac{1}{2}{\rm sgn}\left(A-1\right)\;,
\quad
&\Theta \neq \frac{\pi}{2}({\rm mod} 2\pi), \\[2mm]
0, \quad
&\Theta = \frac{\pi}{2}({\rm mod} 2\pi),
\end{array}
\right\},\;\;
0<F<\frac{1}{2}\;, \\[2mm]
%\end{eqnarray}
%\begin{equation}
& &\sum\limits_{{\rm sgn}(m)}\langle N\rangle=
-{2\over\pi}\arctan\biggl(\tan {\Theta\over2}\biggr)+
\frac{1}{2}{\rm sgn}\left(\tan\frac{\Theta}{2}\right)
\quad,\qquad
\hspace*{3mm} F={1\over2}, \qquad\qquad \\[2mm]
%\end{equation}
%\begin{eqnarray}
& &\sum\limits_{{\rm sgn}(m)}\langle N\rangle=\left\{
\begin{array}{ll}
-\frac{1}{2}
{\rm sgn}\left(A^{-1}+1\right)-& \\[2mm]
-\frac{1}{2}{\rm sgn}\left(A^{-1}-1\right)\;,
\;
&\Theta \neq -\frac{\pi}{2}({\rm mod} 2\pi), \\[2mm]
0, \quad
&\Theta = -\frac{\pi}{2}({\rm mod} 2\pi),
\end{array}
\right\},\;\;
\frac{1}{2}<F<1
\end{eqnarray}
and
\begin{eqnarray}
& &\sum\limits_{{\rm sgn}(m)}
\Phi^{(I)}=-{e^2F(1-F)\over2\pi|m|}\left[{1\over3}\biggl(F-
{1\over2}\biggr)+\right. \nonumber \\
& &\left.+{1\over2\pi}\int\limits_1^\infty{dv\over
v\sqrt{v-1}}{A^{2}v^{2F}-A^{-2}v^{2(1-F)}\over A^{2}v^{2F}
+A^{-2}v^{2(1-F)}+2v-4}\right],
\end{eqnarray}
which are in contrast with the fact that summation of Eq.(3)
over ${\rm sgn}(m)$ yields zero.

Thus, we see that quantum numbers induced  by a singular magnetic
vortex in the fermionic vacuum depend on the gauge invariant
quantities,  $\Phi^{(0)}-\Upsilon$ and $\Theta$. For certain choices
of $\Theta$ the vacuum quantum numbers are periodic in
$\Phi^{(0)}-\Upsilon$ and have definite $C$-parity.

\bigskip

The author would like to thank L.D.~Faddeev, A.V.~Mishchenko,
V.V.~Skalozub and W.I.~Skrypnik for interesting discussions.
The research was supported in part by the State Committee on Science,
Technologies and Industrial Policy of Ukraine and
the American Physical Society.


\begin{thebibliography}{99}
\bibitem{17} It is amusing to note that Eq.(26) has been obtained
by the authors of Ref.[12] at the intermediate stage of
calculations.  However, then these authors add a phantom contribution
to the correct expression and get the incorrect one as a final result;
the latter looks like Eq.(3) which is irrelevant to the case
considered. As to the authors of Ref.[13], who have obtained
also Eq.(26), they, in addition, identify erroneously the limiting
value of this expression at $F \to 1$ with the value of $\langle N\rangle$
at $\Phi^{(0)}-\Upsilon=1$. (Note that the authors of both
Refs.[12,13] consider the case of $\Upsilon=0$).
\end{thebibliography}
\end{document}